\documentstyle[aps,prl,graphicx]{revtex}

\input psfig.sty
\draft
\begin{document}
\twocolumn 
\wideabs{  
\title{Nonequilibrium Effects of Anisotropic Compression
Applied to Vortex Lattices in  Bose-Einstein Condensates}
\author{P. Engels, I. Coddington, P.~C. Haljan, and E.~A. Cornell\cite{qpdNIST}}
\address{JILA, National Institute of Standards and Technology and Department of Physics, \\
University of Colorado, Boulder, Colorado 80309-0440}
\date{\today}

\maketitle

\begin{abstract}
We have studied the dynamics of large vortex lattices in a
dilute-gas Bose-Einstein condensate. While undisturbed lattices
have a regular hexagonal structure, large-amplitude
quadrupolar shape oscillations of the condensate are shown to induce a wealth of
nonequilibrium lattice dynamics. When exciting an m =
-2 mode, we observe shifting of lattice planes, changes of
lattice structure, and sheet-like structures in which individual
vortices appear to have merged. Excitation of an m = +2 mode
dissolves the regular lattice, leading to randomly arranged but
still strictly parallel vortex lines.

\end{abstract}

\pacs{03.75.Fi,67.90.+z,67.40.Vs,32.80.Pj}
} 
\par

The experimental study of vorticity in a dilute-gas Bose-Einstein condensate (BEC) provides
an useful perspective on superfluidity. While the focus of early
work was on the study of single or few vortices
\cite{firstvortex,Madison2000a}, recent advances in technique have made it
possible to create BECs containing large amounts of vorticity
\cite{Madison2001a,Abo-Shaeer2001a,Jilanucleation,Hodby2001a}.
In a rapidly rotating BEC, a lattice of vortices is the lowest energy
state in the rotating frame\cite{Rokhsarnature}. Experimental studies of lattices to
date have concentrated on formation and decay processes
\cite{Madison2000a,Madison2001a,Abo-Shaeer2001a,Jilanucleation,Hodby2001a,Raman2001,Abo-Shaeer2002}.
In this Letter we study the nonequilibrium behavior of vortex
lattices under large-amplitude, anisotropic strain.

\par
\textbf{Experimental technique}
For creating large amounts of vorticity in condensates, we use a
modified version of the spin-up technique described in our
previous paper \cite{Jilanucleation}. The starting point is
a magnetically trapped cloud of $17 \times 10^{6}$  $^{87}$Rb atoms  in the
$|F=1,m_{F}=-1\rangle$ state with a temperature approximately three times above
the critical temperature for the BEC phase transition,
$T_{c}=67$ nK. By manipulating the ellipticity of the confining potential, we can resonantly pump angular momentum into the as yet uncondensed cloud \cite{spinup}.
We then tune the trapping potential to near perfect axial symmetry with 
trap frequencies $\{\omega_{\rho},\omega_{z}\}=2\pi\{8.35,5.45\}$Hz. In this
round trap the cloud continues to rotate, and by performing a
nearly one-dimensional evaporation along the axis of rotation (z-axis) as
described in Ref. \cite{Jilanucleation}, we cool down and further
spin up the cloud until a highly rotating condensate is born out
of the rotating normal cloud. For the experiments described in
this paper, the evaporation is continued to a point where little or no
 thermal fraction remains. As a result of this procedure,
we obtain highly rotating condensates with typically
$1\times10^{6}$ atoms, containing 130 vortices or more.

To study these condensates we can image along the x, y, or
z directions. When the condensate is held in the trap, the core
size of the vortices is close to the resolution limit of our
optical detection system so that individual vortices are not
resolved by in-trap imaging. However, by nondestructively imaging a trapped condensate from the side (along the x-direction) and measuring
its aspect ratio, the rotation rate of the condensate
can be determined \cite{Jilanucleation,Raman2001}. For our
highly rotating condensates we determine typical aspect ratios
of 0.5, which is markedly different from the static aspect ratio
of the magnetic trap, 1.53, because of centrifugal forces.  Typical condensate rotation rates of $0.95
\omega_{\rho}$ are inferred.

\begin{figure}
\begin{center}
\psfig{figure=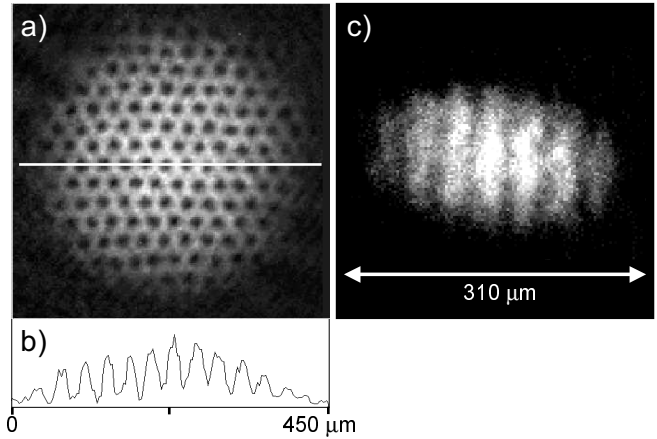,width=1\linewidth,clip=}
\end{center}
\caption {(a) Expansion picture of a vortex lattice seen along the rotation axis.
(b) One pixel wide cross section along the white line in (a).
(c) Expansion picture of a different condensate rotating more slowly than
in (a), seen from the side. }
\label{latticeexpansion}
\end{figure}

To resolve individual vortices, we can release a
condensate from the trap and image it after it has expanded by a factor of 5 \cite{Madison2000a,vortexrings}. Such expansion pictures taken along the
z-direction reveal a hexagonal lattice of vortex cores in highly
rotating condensates [Fig.~\ref{latticeexpansion}(a)].
The lattice structure is remarkably regular even at the outer
regions of the condensate, and is observed with very good
contrast as shown by the cross section in
Fig.~\ref{latticeexpansion}(b).

When looking at such vortex lattices in expansion from the side
(i.e., along the x or y direction), good contrast is only
obtained if the lattice is oriented in such a way that vortices
line up behind each other along the direction of view, or in
other words when the direction of view is parallel to lattice
planes. The vortex lattices are not stationary in the lab frame, but are
rotating with the condensate, and furthermore the initial
orientation of the lattice in the xy plane cannot be controlled
in current experiments. Therefore the times when vortices line
up along the direction of sight are unpredictable, and images
revealing lattice planes are only obtained in a random subset of
all trials. One example is shown in
Fig.~\ref{latticeexpansion}(c), where the vortex planes are
clearly visible as dark vertical lines.
These images are direct verifications of the theoretical prediction
that vortex lines are only marginally bent for our experimental parameters
\cite{Garcia,Feder}, as is also confirmed by the good contrast seen in
the topview picture of Fig.~\ref{latticeexpansion}(a).

\par
Vortex lattices such as the one shown in Fig.~\ref{latticeexpansion} are the
starting point for our studies of the effects of dynamically
generated strain. We apply the distortion by
resonantly driving quadrupolar shape modes in the vortex-filled
condensate \cite{Chevy2000,Tiltmodes}.  In the large-amplitude
limit, the $m_{z}=+2$ and $m_{z}=-2$ modes distort the circular
cross-section of the condensate into an ellipse and cause the
major axis of the ellipse to rotate along with, or against, the
sense of the condensate rotation, respectively. The generated
strain is considerable -- the ratio of major to minor axis of
the condensate can be larger than three to one.  The frequencies
of the $m_{z}=\pm2$ modes may be calculated using the sum rule
argument given by Refs.\cite{ZambelliStringari1,ZambelliStringari2}. Our condensates are rotating
typically at $0.95 \omega_{\rho}$, just 5\% under the centrifugal limit.
At the centrifugal limit, the major axis of the $m_{z}=+2$ ellipse is
fixed in the rotating condensate frame, and the $m_{z}=-2$ shape is
fixed in the lab frame.  At our rotation rate of $0.95 \omega_{\rho}$,
the distortion, as seen in the frame of the lattice, rotates at
only 0.4 Hz for the $m_{z}=+2$ mode, compared with 8.34 Hz for the $m_{z}=-2$
mode. For this reason \cite{fluidflowpattern}, we expect very different
resulting lattice dynamics in the presence of either of the two modes.

\par
\textbf{Lattice dynamics in the presence of an $m_{z}=-2$ surface
mode} Because the $m_{z}=-2$ mode is almost stationary in the lab frame for the parameters of our experiment,
this mode can be excited conveniently and
nearly resonantly by a static trap deformation. Indeed we
observe that by jumping from a round trap to a trap that has an
ellipticity of 3.6\% \cite{defellipticity}, the ellipticity of the condensate
increases over a timescale of 300 ms from approximately 0 to
40\%, thereby exceeding the trap ellipticity by more than an
order of magnitude. Since the vortex lattice is rotating quickly
with respect to this almost static deformation, it is an
interesting and nontrivial question to ask whether or not a
lattice structure is maintained, and, if so, how it rearranges
in the presence of the excitation.

\begin{figure}
\begin{center}
\psfig{figure=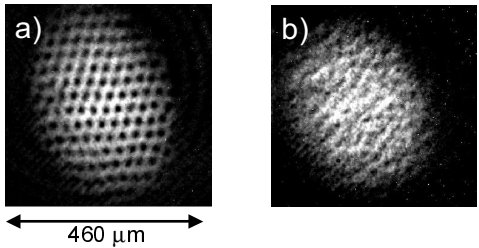,width=1\linewidth,clip=}
\end{center}
\caption {Lattice evolution after an $m_{z}=-2$ excitation. Pictures
taken (a) 173~ms and (b) 873~ms after start
of trap deformation}.
\label{globaltimesequence}
\end{figure}

\begin{figure}
\begin{center}
\psfig{figure=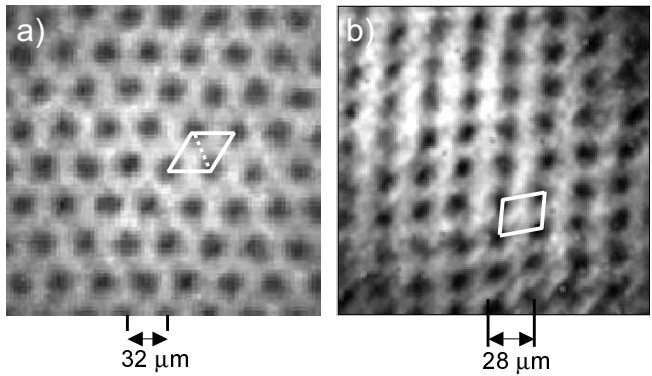,width=1\linewidth,clip=}
\end{center}
\caption {Change of lattice structure.
(a) Hexagonal structure in an undisturbed lattice. (b)
Near orthorhombic structure seen transiently during lattice evolution in the
presence of an $m_{z}=-2$ quadrupolar surface mode. }
\label{hexaortho}
\end{figure}

Experimentally, vortex lattices can indeed be observed in
expansion images  up to 400~ms after the start of the continuously applied
trap deformation (Fig.~\ref{globaltimesequence}). 
Subsequently, long range order is lost and vortex visibility becomes low, presumably due to tilting \cite{Tiltmodes} or bending of the vortex lines \cite{Abo-Shaeer2001a}.
 A closer look at the first 400~ms of
this evolution reveals a wealth of intriguing vortex dynamics.
As the vortex lattice rotates in the deformed condensate, the
lattice planes must continuously shift relative to each other to
accommodate to the elliptical shape of the condensate. One
consequence of this is transient changes of the lattice structure.
In Fig.~\ref{hexaortho} the lattice has changed from the hexagonal
structure of an undisturbed lattice to a near orthorhombic
structure. 
\par
\begin{figure}
\begin{center}
\psfig{figure=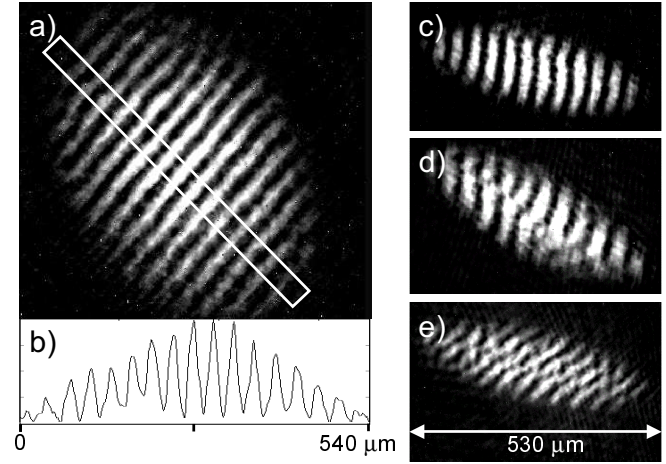,width=1\linewidth,clip=}
\end{center}
\caption {
(a) Sheet-like structure as seen during lattice evolution in the
presence of an $m_{z}=-2$ quadrupolar surface mode. (b) Cross section integrated over the white box in (a). Even though the box is wider than the calculated vortex core spacing, the observed contrast is nearly perfect.
(c-e) Same as (a), but observed in a more deformed trap with
$\{\omega_{x,y,z}\}=2\pi\{6.0,8.6,13.8\}$Hz }.
\label{sheets}
\end{figure}

The most striking observation, however, is of
condensates containing sheet-like structures rather than
individual vortex cores, as shown in
Fig.~\ref{sheets}(a). We interpret these sheets as rows of
vorticity along which
individual vortex cores have essentially merged. Given the near perfect contrast of
these sheets, as shown by the cross section in
Fig.~\ref{sheets}(b), we can exclude the possibility that the sheets are
merely formed by a collective tilting of vortices along a
lattice plane, which would result in an intermediate contrast.
Similarly, we can rule out that this is an effect merely of
imaging resolution - had the vortices retained their original
structure but come so close together that we could not spatially
resolve them, we again would see white stripes set off by grey
troughs, rather than the near 100\% alternating stripes of white
and black.  If there continue to exist wispy fingers of condensate
that cross the stripes and differentiate individual vortices, they
must be very tenuous indeed to be consistent with the observed
contrast. The clouds have been continuously distorted from their
original hexagonal symmetry, so the observation of 16 stripes
along the major axis of the BEC in Fig.~\ref{sheets}(a) tells us that there
must also be the equivalent of 16 units of vorticity along the
length of the stripes through the center of the cloud. By measuring the in-trap height of the condensate, the ellipticity in the xy-plane and the number of atoms, we can infer the ratio of unit vorticity spacing (along the minor axis) to healing length (given by 
$\frac{1}{\sqrt{8\pi na}}$, where n is the mean density and a the scattering length) to be on the order of 5.

\par
\begin{figure}
\begin{center}
\psfig{figure=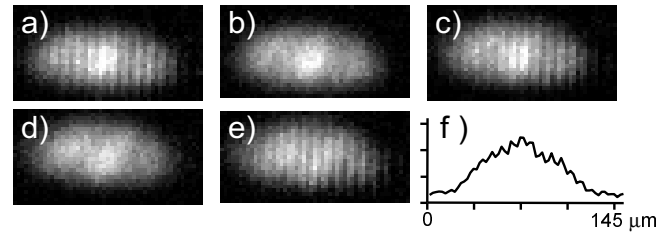,width=1\linewidth,clip=}
\end{center}
\caption {(a-e) Nondestructive in-trap images of the sheet-like structures seen along the x-direction [conditions similar to Fig.~\ref{sheets}(a)].
Spacing between images 10.6 ms. Note the very different spatial scale from expansion images (e.g., Fig.~\ref{latticeexpansion}). (f) Cross section of (a),
integrated over the condensate.}
\label{lighthouse}
\end{figure}

In fact the contrast of this sheet-like phase is so deep that
the structure can be detected by looking at condensates \textit{in trap}
(along the x or z direction), even though the structure
is close to our optical resolution limit. By taking a time series
of ten nondestructive images of a single condensate
along the x direction, we see that the sheet structure appears
and disappears periodically with a period of about 21.2 ms for a
condensate rotating at a frequency of $0.95
\omega_{\rho}$ [Fig.~\ref{lighthouse}]. 

This periodicity can be understood as follows: The sheet structure
forms when a lattice vector lies along a minor axis of the cloud,
presumably since this is when the vortices are closest together.
As that vector rotates past the minor axis, structure along the vector is reestablished, the two-dimensional (2D) pattern of individual vortices reemerges, and the high-contrast stripes disappear from the in-trap images. A lattice vector aligns with the minor axis six times per lattice rotation period due to the sixfold symmetry of the unperturbed lattice, consistent with the observed frequency of periodic stripe formation. This interpretation is bolstered by the expansion images, which show high contrast sheets when a lattice vector lies close to the minor axis, but 2D lattices (albeit distorted) for other orientations. We find it remarkable that this lattice-to-sheet-to-lattice sequence persists through multiple cycles before long range order and visibility are compromised.

Using a trap with a stronger distortion of 34\% as in Fig.~\ref{sheets}(c-e), we also observe sheets breaking apart in the center.
In Fig.~\ref{sheets}(c), the density distribution along the lattice vector
that we see near-parallel with the minor axis is nearly
featureless. Figure~\ref{sheets}(d,e) hint at the succeeding lattice evolution: Fig.~\ref{sheets}(d) represents a small rotation from Fig.~\ref{sheets}(c), and
we see that the sheets have begun to reconnect across the cloud.
Figure~\ref{sheets}(e) appears to be a continuation of that process, with considerable evidence of uniformity in the repeating structure.

\par
\textbf{Lattice dynamics in the presence of an $m_{z}=+2$ surface
mode} We excite the $m_{z}=+2$ mode by elliptically deforming the magnetic
trap and rotating that ellipse around in the xy plane. 
When the $m_{z}=+2$ mode is driven approximately 1~Hz away from its resonance, the ellipticity of the condensate increases from 0 to 50\% over a timescale of 250~msec. Under the influence of an on-resonant drive, the ellipticity increases continuously, even passing beyond 90\% after 700~msec and approaching 100\% \cite{Rosenbusch}. As discussed above, the distortion rotates only very slowly with
respect to the lattice, and thus we expect it to be 
correspondingly ``gentle" with respect to its action on the
lattice.

\begin{figure}
\begin{center}
\psfig{figure=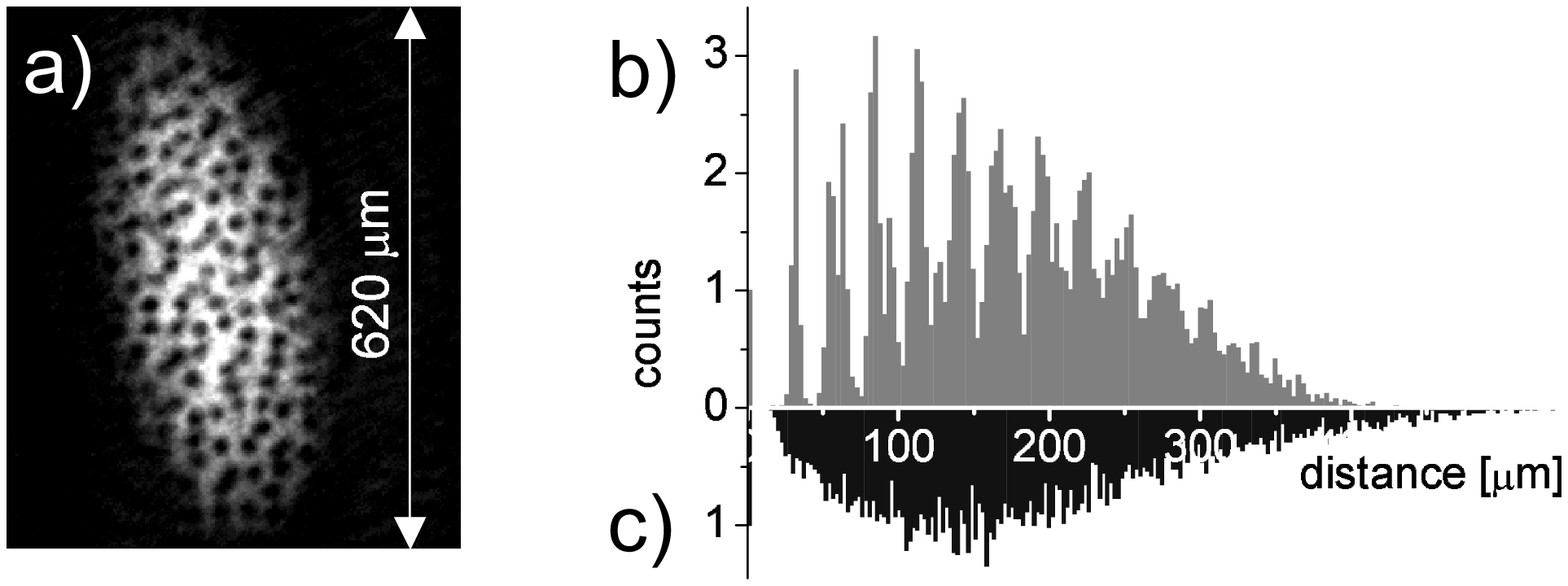,width=1\linewidth,clip=}
\end{center}
\caption {(a) Irregular assembly of vortex cores as observed after 500~ms in the presence of an $m_{z}=+2$ quadrupolar surface mode. (b) Histogram of distances between each pair of vortices in the regular hexagonal lattice shown in Fig.~\ref{latticeexpansion}a. The visibility of separate peaks reveals the high degree of long-range order. (c) Same as (b) but calculated for the BEC of Fig.~\ref{liquidphase}a. Peaks are barely visible here, meaning that the lattice has almost reached a random disorder.}
\label{liquidphase}
\end{figure}

Indeed, as we show in Fig.~\ref{liquidphase}(a), we do not see the lattices
squeezed into sheets for the $m_{z}=+2$ mode. 
Instead we observe a gradual increase of disorder in the lattice. 400~msec after we start the excitation, we see a complete loss of long-range order
in the vortex structure. The imaging contrast of individual
vortices remains high, so we know they remain parallel to the line of sight (in contrast for instance to the case of Fig.~\ref{globaltimesequence}), but
otherwise they appear to have liquified (see the correlation
histogram of Fig.~\ref{liquidphase}).
This behavior confirms our expectation of the $m_{z}=+2$ mode being a gentler disturbance for the lattice than the $m_{z}=-2$ mode.
\par
\textbf{Conclusions}
In conclusion we have succeeded in observing 
a wealth of nonequilibrium vortex lattice dynamics in a BEC under anisotropic compression. Key observations include the shifting of lattice planes relative to each other, leading to transient changes of the lattice structure, the formation of an unexpected and very pronounced sheet structure, and finally, a gradual transition from an ordered lattice to a rather liquid-like arrangement in which vortex  cores remain visible, but their relative positions are irregular.
The observed sheet structure is of
particular interest: at present the
behavior of densely packed vortex lattices is in the center of
theoretical discussions about vortices \cite{Ho2001,baym,cirac}.
While most of these theoretical discussions propose to enter
this regime by having  high
rotation rates almost inaccessible to experiment or inconveniently low atom numbers in the condensate, this paper demonstrates an
alternative and unexpected way of bringing vortices close
together. We hope that the results presented in this Letter
will stimulate further theoretical discussions in the
field of vortex lattices.

\par
The work presented in this paper was funded by NSF and NIST. PE
acknowledges support by the Alexander von Humboldt Foundation.


%

\begin{thebibliography}{10}
\bibitem{fluidflowpattern}
We emphasize as well that the two different modes
result in very different fluid-flow patterns. The $m_{z}=+2$ mode very
closely approximates rigid-body rotation, while much of the $m_{z}=-2$
velocity field is provided by an irrotational, quadrupole flow
pattern. This distinction may provide a parallel explanation for
the correspondingly distinct observed effects.
\bibitem{defellipticity}
We define ellipticity as $\frac{\omega_{x}^2-\omega_{y}^2}
{\omega_{x}^2+\omega_{y}^2}$ with $\omega_{x}$ and $\omega_{y}$ being the trap
frequency along the minor and major axes of the ellipse, repectively.
\end{thebibliography}
\end{document}